# Telepresence Interaction by Touching Live Video Images

JIA YUNDE, XU BIN, SHEN JIAJUN, PEI MINTAO, DONG ZHEN, HOU JINGYI, and YANG MIN, School of Computer Science, Beijing Institute of Technology, CHINA

Beijing Lab of Intelligent Information Technology, CHINA

This paper presents a telepresence interaction framework based on touchscreen and telepresence-robot technologies. The core of the framework is a new user interface, Touchable live video Image based User Interface, called TIUI. The TIUI allows a remote operator to not just drive the telepresence robot but operate and interact with real objects by touching their live video images on a pad with finger touch gestures. We implemented a telepresence interaction system which is composed of a telepresence robot and tele-interactive objects located in a local space, the TIUI of a pad located in a remote space, and the wireless networks connecting the two spaces. Our system can be a perfect embodiment of a remote operator to do most of daily living tasks, such as opening a door, drawing a curtain, pushing a wheelchair, and other like tasks. The evaluation and demonstration results show the effectiveness and promising applications of our system.

## 1. INTRODUCTION

As Wireless internet and mobile communication networks are becoming ubiquitous, we can easily establish our presence in a remote location through telepresence technologies, e.g. using a pad or smartphone to video chat with family members, watching a kid's game over a tablet, and holding a party in a virtual room [Rae et al. 2015a]. The more recent emergence of telepresence robots or robotic telepresence systems [Kristoffersson et al. 2013; Suitable Tech. 2016] can provide a more flexible telepresence experience by allowing an operator to have some degree of mobility in a remote location. The existing telepresence technologies have shown significant efforts on the verbal and non-verbal tele-communication [Heath and Luff, 1991; Rae et al.

This work is supported by the National Science Foundation of China, under Grants No.61472038 and No.61375044.

Authors' address: 5 South Zhongguancun Street, School of Computer Science, Beijing Institute of Technology, Beijing, 100081, China.

Correspondence should be addressed to: JIA Yunde; Email: jiayunde@bit.edu.cn

2013, 2015b], they are insufficient in telepresence for teleinteraction and teleoperation. The teleoperation and teleinteraction are necessary for users to transcend the space in order to save commuting times, stay out of harm's way, and get unprecedented scope in collaboration worldwide [Sha and Agrawala 2006]. The simplest way to realize the telepresence for teleoperation and teleinteraction is to modify a humanoid robot [e.g., Meeussen et al. 2010; Banerjee et al. 2015; Philips et al. 2016] to be your embodiment. The operator can pilot such a humanoid robot to perform any interaction and operation tasks in a remote space to replace his/her direct contacts. However, a humanoid robot is too expensive and complex for daily living applications. How a telepresence robot could provide both moving and operating experiences for a remote operator is an interesting topic.

We have observed that when we use touch-screen technologies to realize the teleoperation of an object (e.g. a toy car), we have got accustomed to the guidance of pressing graphic buttons beside or overlapping the video window [Micire et al. 2011; Tsui et al. 2015]; We also tried to directly touch and drag the live video image of the toy car, activated by our touching impulse [Fleischhut 2011], and hoped to see its action response (such as stop or turn round). In this work, we focus on telepresence interaction with physical objects by directly touching their live video images fed by a telepresence robot. The telepresence interaction includes the verbal and non-verbal communication, teleinteraction, and teleoperation, and thus remote operators' experiences would be so similar to actually being there that there would be no noticeable difference as envisioned by [Minsky 1980]. Our aim is to extend telepresence interaction into the lives of ordinary people.

This paper presents a telepresence interaction framework based on touchscreen and telepresence robot technologies. The core of the framework is a new user interface, **T**ouchable live video **I**mage based **U**ser **I**nterface (TIUI). The TIUI allows a remote operator to not just drive the telepresence robot, but also operate and interact with physical objects only by touching their live video images on the touchscreen of a mobile device, such as a pad or smartphone. In other words, the TIUI on a pad enables an operator in a remote space to readily access the system to drive the telepresence robot and perform many tasks as if he/she did it in the local environment (space). Figure 1 illustrates an example task of the telepresence interaction system, where the telepresence robot is capturing the live video of the

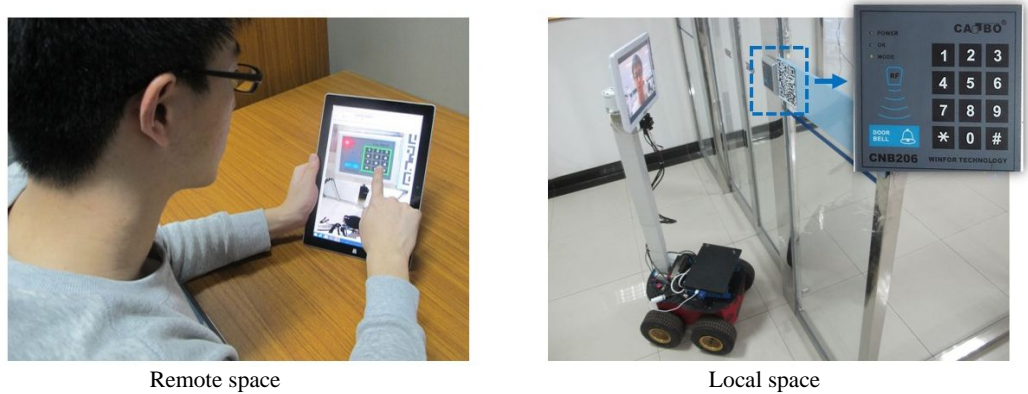

Remote space          Local space

Fig. 1. Illustration of the telepresence interaction by touching live video images. Right: the telepresence robot located in a local space is capturing the live video of the password access control panel of an auto-door. Left: a remote operator is pressing the live image of a button via the TIUI on a pad to open the door. (Online demo at http://iitlab.bit.edu.cn/mcislab/projectdetail.php?id=12)

password access control panel of an auto-door in the local space, and a remote operator is pressing the live video image of a button by using the TIUI on a pad to open the door (See demo at http://iitlab.bit.edu.cn/mcislab/projectdetail.php?id=12). A remote operator is able to not just drive such a telepresence robot but operate any local objects, such as doors, elevators, appliances or office settings by touching their live video images.

The characteristics of the telepresence interaction system are described as follows,

- Activeness: A remote operator can use the telepresence robot to actively capture the live video of a local environment for telepresence interaction;
- Pervasiveness: A remote operator can readily access the system anywhere through wireless internet by using the TIUI on a pad to perform telepresence interaction;
- Feelings of presence: A remote operator can use the TIUI to operate physical objects of a local space by touching their live video images as if he/she touched them in the local space.

The remainder of this paper is organized as follows. Section 2 surveys related work in live video-based teleoperation and telepresence robots. The system architecture for telepresence interaction by touching live video images is proposed in Section 3. Section 4 describes the telepresence robot developed in our lab for testing the framework of telepresence interaction. Section 5 proposes a new user interface, the TIUI, and its function designing. Section 6 presents the software components of the system and the system state transition. Section 7 reports the evaluation the usability of the system, including driving the telepresence robot and interacting with local physical objects only by touching live video images on the TIUI of a pad. Section 8 gives the demonstrations of the system in driving the telepresence robot, interacting with local physical objects, and pushing a wheelchair. Section 9 discusses the technological challenges, the acceptability, limitations, and the future work of the system, and we conclude this work in Section 10

We must say that this work covers so many technical details, but we are just able to give the high level description and a few of the requisite technical details due to the space limitation. The rest will be published elsewhere.

## 2. RELATED WORK

A great deal of work related to telepresence technologies has been done over the last decades, ranging from the basic of videoconferencing setups to humanlike androids, and used to support many specific domains, such as medical [Wechsler et al. 2013; Marini et al. 2015], education [Tanaka et al. 2013; Meyer 2015], and office settings [Venolia et al. 2010; Procyk et al. 2014; Neustaedter et al. 2016]. As our work focuses on the telepresence interaction based on touchscreen and robotic telepresence technologies, we will not review literature related to general telepresence, but only consider live video-based teleoperation and telepresence robots.

### 2.1 Live Video-Based Interaction

An early attempt to interact with physical objects in live videos stems from [Tani et al. 1992]. They explored 2D and 3D models to position physical objects in the live video and used a mouse and keyboard based user interface to realize teleoperation and teleinteraction in a structured environment. They implemented a system called HyperPlant to remotely control an electric power plant. We use a new user interface (TIUI) to directly touch the live video images of physical objects with finger touch gestures to perform teleoperation and teleinteraction tasks in an unstructured environment without modelling objects. In recent years, there is some work reported in the literature about touchscreen technologies used to manipulate live video images for teleinteraction. TouchMe [Hashimo et al. 2011] adopted a touch-screen based user interface with the live video acquired from a ceiling camera to manipulate a complex multi-DOF robot in a remote location, and they used a computer graphic model of the robot which overlays its real image synchronously to help the user predict how the robot will move. CRISTAL [Seifried et al. 2009] designed an interface on a multi-touch tabletop surface with the live video from a ceiling camera to control daily living home devices in the same room. These systems require complex geometric models for the positioning and mapping of manipulated objects. Moreover, these systems use a ceiling camera configuration as the third-person view to offer a whole picture of the environment so that all participants are visible, but they are the most likely to suffer from occlusion problems.

Using a ceiling camera as the third-person view, some intuitive teleoperation strategies without object or robot models have been explored, such as sketching the behaviors and intended paths of an indoor robot on a touchscreen [Sakamoto et al. 2009; Liu et al. 2011; Frank and Kapila 2015], controlling multiple mobile robots simultaneously by manipulating a vector field on the live video image [Kato et al. 2009], interacting with multiple robots using toys on a large tabletop display [Guo et al. 2009]. These systems were designed and evaluated for the situation in which operators and robots are located in the same space or environment, and operators can see the physical objects in person to monitor them. In contrast, our system only uses the live video image as visual feedback for teleoperation and teleinteraction.

There is an increasing interest in using pads, smartphones, or tablets to capture live videos as the first-person view to interact with physical objects in the same location. [Boring et al. 2010, 2011] presented a system called Touch-Projector that enables a user to interact with remote screens through a live video image on their mobile device. exTouch [Kasahara et al. 2013] is a wonderful interaction system which enables a user to control the physical motion of an actuated target by simply touching and dragging through an augmented reality mediated live video interface on the touchscreen of a pad. Our system also uses live video based interface which allows an operator to interact with physical objects in a remote environment based on telepresence robot technologies.

**2.2 Telepresence Robots**

A telepresence robot is characterized by a video conferencing system mounted on a mobile robot base, which allows a remote operator to move around and communicate with people in the robot's environment. So far there is much work reported in literature about telepresence robots and their applications. Interested readers are referred to [Kristofferson et al. 2013] for a good comprehensive literature survey.

The earliest telepresence robot was reported by [Paulo and Canny 1998]. Its basic architecture includes a mobile robot base, an LCD screen, a color video camera, microphone, speaker, a robot hand/arm hardware with a 2 DOF pointer and a laser pointer attached to its tip for simple gesturing. A pilot in a remote space can tele-operate the robot as well as pan/ tilt the head by using a keyboard and a joystick. Following this work, many types of telepresence robots have been assembled, and some of them are commercially-available, such as Giraff [Giraff 2016], QB [Anybots 2016], Texai [Willow Garage 2016], VGo [VGo 2016], and Jazz [Gostai 2016]. The mainstream of the existing systems share the common design and functionality, i.e. a mobile robot base attached with a human-height pole, and audio/video communication devices mounted on the pole. The Giraff robot comes with a tilt-standing screen mounted on a non-adjustable pole attached to a mobile robot base. It uses a web-cam with a wide-angle view for the pilot, and is designed for aging-in-place in home environment. The robotic platform is accessed and controlled via a standard computer/laptop using a software application. The QB robot employs two cameras: one non-tilt forward facing camera and one downward facing camera for navigation. It can be piloted from any browser on Windows and Macintosh and is suited for office environments. The Texai robot also can be piloted from standard computer/laptop by using a mouse, and contains a screen, a six-microphone array enabling pilot users to localize directions of sound, and two wide-angle HD cameras, one forward facing and one downward facing. The VGo robot can be piloted from both Windows and Macintosh and navigation is done by "click and go" buttons and arrow keys. The Jazz robot uses smartphone technologies, which allows a user to access the robot system from smartphones or computers via a web-based interface, and the user can simply click on the real-time image displayed on the interface by using the 3D pointer to indicate the direction to follow. All the existing robotic telepresence systems have faced challenging issues related to navigation, interaction, and accessibility [Neustaedter et al. 2016]. In order for greatly alleviating these issues, we developed a telepresence robot in our lab for telepresence interaction, particularly, it can be driven by directly touching its live video on the touchscreen of a pad with finger touch gestures.

## 3. SYSTEM ARCHITECTURE

Figure 2 shows the system architecture for telepresence interaction by touching live video images. The system is composed of a telepresence robot and tele-interactive objects located in a local space, the TIUI on a pad located in a remote space, and the wireless internet communication network connecting the two spaces.

A tele-interactive object refers to a physical object having three attributes: identification, actuation, and wireless internet communication. In practice, we can design a device with the three attributes as a standard box to be added on a physical object to make the object tele-interactive. In the following description of this paper, for simplicity, we use the term object that is identical with tele-interactive object or tele-interactive device. With the increasing demand for smart buildings and smart homes, many tele-interactive objects or devices are emerging, and many office settings and housing appliances are becoming tele-interactive for teleoperation and teleinteraction through smart phones. The identification (ID) data of an object contains its name, location, operation interface, operation command, and driver software. The term actuation refers to applying power to a relevant component to realize its specific function, such as moving, heating, expanding, presenting, etc. The wireless internet communication enables an operator to access the internet through mobile devices. In our system, WiFi internet access is used to realize wireless communication, which connects a remote space with a local space for telepresence interaction by touching live video images.

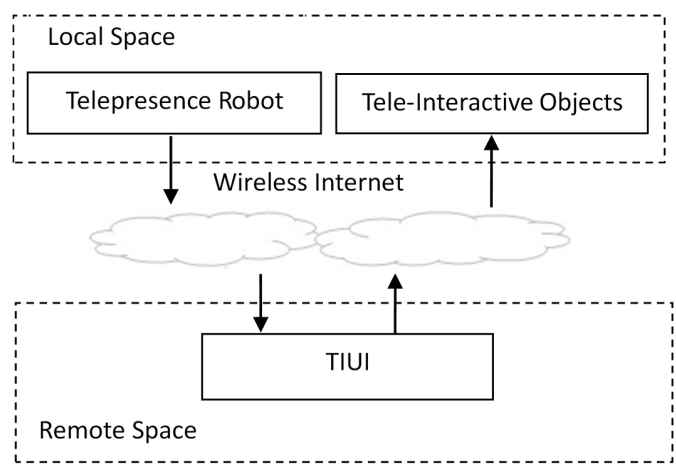

Fig. 2. System architecture

A telepresence robot, in general speaking, is one of the family members of tele-interactive objects as it does have the three attributes for teleoperation. A telepresence robot is regarded as the embodiment of a remote operator to perform almost any interaction and operation tasks he/she likes to do. Therefore, for a clear presentation, we exclude the telepresence robot from tele-interactive objects. The telepresence robot can be piloted by a remote operator to actively capture the live video of a local environment for telepresence interaction, which has resolved the problems of occlusion and limited view field.

The TIUI, Touchable live video Image based User Interface, is the core of the telepresence interaction system. The TIUI is just a plain live video window on the touchscreen, which allows a remote operator to drive the telepresence robot and operate local physical objects by directly touching their live video images with finger touch gestures, instead of clicking the buttons and arrow keys like the traditional GUI. By using the TIUI, one can directly touch the live video images of any existing control panels with good ergonomics in a remote space as if he/she were there to touch the physical control panels. So it is obvious that touching live video image of objects creates much better feelings of being present than touching graphical buttons and/or arrow keys.

Figure 3 shows a smart home, as a test bed of the local space, built in our lab for evaluation of our system. The smart home contains tele-interactive objects including an auto-door with password access control (A), a lighting tree (B), a telepresence robot (C), an electric curtain (D), and an electric wheelchair (E).

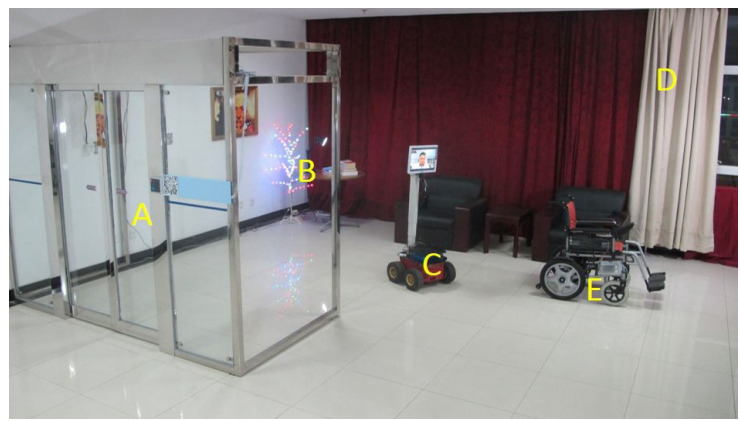

Fig. 3. Illustration of a local space simulating a smart home for testing our system. The smart home contains tele-interactive objects, including an auto-door with password access control (A), a lighting tree (B), a telepresence robot (C), an electric curtain (D), and an electric wheelchair(E).

## 4. TELEPRESENCE ROBOT

The existing commercial telepresence robot are designed for teleoperation by using the GUI with a mouse, joystick, buttons, and/or arrow keys, and cannot be used to realize our system to perform telepresence interaction by touching live video images. We developed a telepresence robot, called the Mcisbot, which be remotely operated by touching its live video image. We then designed a new user interface TIUI for a remote operator to pilot the robot and interact with local physical objects. The TIUI design will be described in Section 5.

The Mcisbot robot follows the basic structure and functionality of telepresence robots [Paulo and Canny 1998; Willow Garage 2016] which contains a mobile robot base and a telepresence head, as shown in Figure 4. W used the Pioneer 3-AT on our hand as the mobile robot base, and specially designed a proper robot head for testing the functionality of the system. The robot head consists of a light LCD screen, a forward-facing camera (FF camera), a downward-facing camera (DF camera), and a speaker & microphone, and all together are mounted on a pan-tilt platform hold up by a vertical post. The post can be moved vertically to adjust the robot height ranging from 1200mm-1750mm, covering school child and adult body heights. All components of the head are inexpensive and commercial off-the-shelf products. Although we used the costly Pioneer 3-AT on our hand as the mobile robot base of the Mcisbot, we can equip any other mobile robots or vehicles with our robot head to assemble a very low cost telepresence robot [Shen et al. 2016].

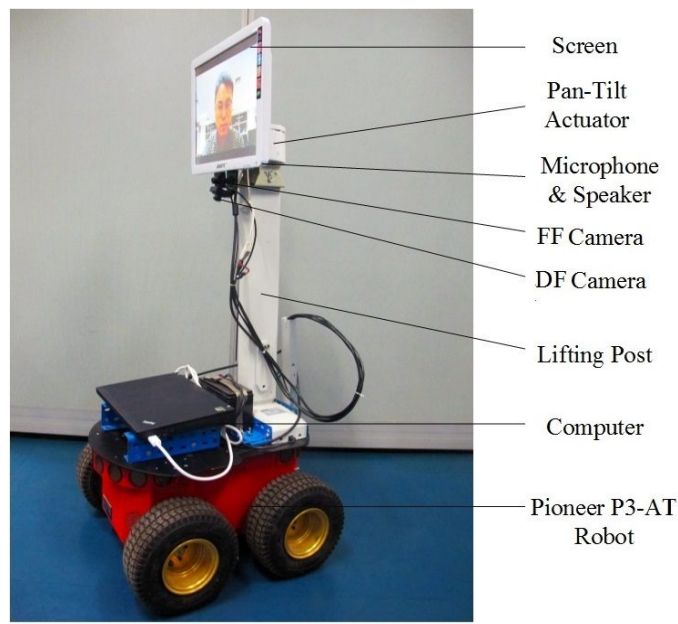

Fig. 4. A picture of the Mcisbot robot

Some of the existing telepresence robots are equipped with two cameras: one camera (FF camera) for video communication and the other camera (DF camera) for navigation [Lee and Takayama 2011; Neustaedter et al. 2016]. The two cameras can provide two live videos displayed on two windows, respectively, for visual feedback. Our Mcisbot robot is also equipped with the FF camera and the DF camera for both visual feedback and live video manipulation. The FF camera with a high resolution is used to capture the live video image (FF image) of the front environment for a remote operator to look forward or/and select targets to be operated. The FF image is also used to perform recognition task for telepresence interaction. The DF camera with a fish-eye lens is used to capture the live video image (DF image) of the ground around the robot for a remote operator to drive the robot, and also for the robot to execute semi-autonomous navigation. However, in testing visual feedback with these two video windows, we found that two windows would introduce some confusion over the local space. For example, a remote operator felt missing some views and the context of the local space, and distracted by switching attention between the two windows and adapt the different windows. Fortunately, there is a great deal of overlap between the two live video images owing to the fish-eye lens of the DF camera, and we can easily stitch these two live video images as one image to display in one video window. We call the stitched image the forward-downward-facing image, or FDF image. Figure 5 shows an example of a stitched image (FDF image) created by the two live video images from the FF camera and the DF camera, respectively.

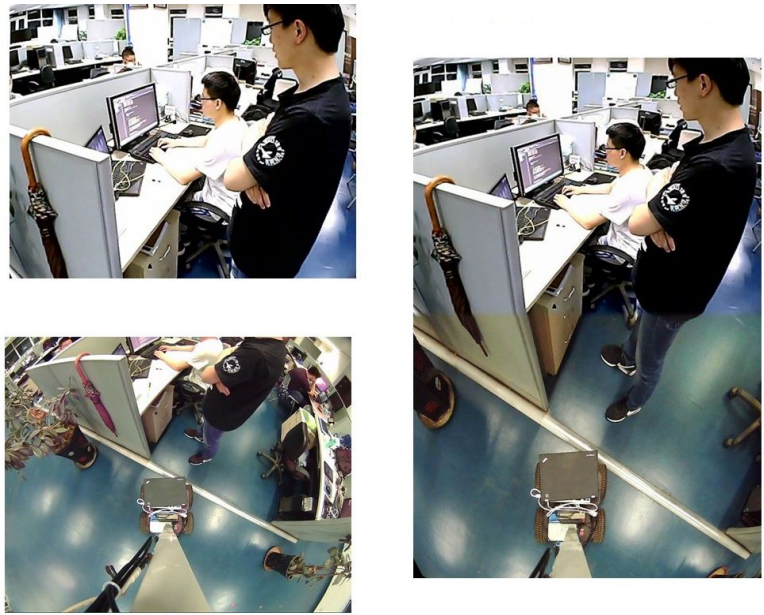

Fig. 5. Illustration of a stitched video image. Left top: a live video image from the FF camera. Left bottom: a live video image from the DF camera. Right: the stitched video image for the TIUI which is formed by stitching the two live video images from the FF camera and the DF camera, respectively.

From Figure 5 we can see that the intensities of the upper and lower parts of an FDF image are apparently different as the two cameras are off-the-shelf products and have their own automatic gain controls. We did not smooth this difference for reserving the FF image quality and saving the computational cost as well. In our practical test, this difference is no hurdle for operators to perceive remote environments.

The FDF live video image provides a required large view of a local environment in the TIUI, which makes an operator look forward clearly to recognize objects and look downward naturally with a very large view of field for navigation, as if he/she was being there to look forward to capture targets and look down to move safely. [Tsui et al. 2014, 2015] designed a camera-hat of the telepresence robot Vgo containing three cameras with regular lenses and stitched the individual output of the three cameras together to form a vertical panoramic video stream for a wider field of view. But this camera-hat neither captured panoramic image from the forward-facing view of the front targets to the downward facing view of the mobile robot base, nor created a high quality live video stream for telepresence interaction by touching live video images.

It is the Mcisbot robot, as the embodiment of a remote operator, to make the telepresence interaction system have the significant property of activeness. The Mcisbot robot can provide a useful mediation for a remote operator to move around in the local space and actively capture the local live video images for telepresence interaction.

## 5. TOUCHABLE LIVE VIDEO IMAGE BASED USER INTERFACE

The rapid development of smart mobile technologies has provided important tools for communication, and changed how we work, live, communicate, and interact socially. Currently, remote people can readily watch what objects are there in offices, homes, or any other places via smartphones or pads. Let us imagine if we can remotely operate or interact with physical objects by touching their live video images while watching them, how wonderful it will be. To this end, we propose a new user interface, **T**ouchable live video **I**mage based **U**ser **I**nterface (TIUI), by using touch-screen and computer vision technologies. The TIUI enables an operator to touch the live video images of physical objects on the touchscreen in a remote environment to operate and interact with them.

### 5.1 Interface Functions

The conventional touchscreen based user interface for teleoperation of a robot often contains one or more live videos as visual feedback and graphical buttons to control the robot [Micire et al. 2009; Tsui et al. 2015]. In contrast, our TIUI is a plain FDF live video fed by the Mcisbot robot, which is not only acting as visual feedback and videoconferencing, but also carrying a great number of interaction functions just like the physical objects. The interaction functions of a live video are often related to perceived affordance of physical objects [Gibson 1979; Norman 2008]. When we see an image of a door on the screen, for example, we can touch its knob image to open it, or push its button image right by its side to open it. Generally, the TIUI enables a remote operator to directly touch objects' images on the touchscreen with finger touch gestures to realize teleoperation and teleinteraction in more natural and efficient way, i.e., the remote operators' experience would be so similar to actually being there.

Designing the TIUI has to resolve two issues: one is to capture and interpret the finger touch gestures corresponding to interaction commands, and the other is to recognize, localize, and track the touched object in a live video. Specifically, we expected to design finger touch gestures for the TIUI to perform the three types of teleoperation and teleinteraction tasks:

(1) One can touch the live video image of a telepresence robot with finger touch gestures to pilot the robot;
(2) One can touch the live video image of physical objects with finger touch gestures to operate and interact with them;
(3) One can touch the live video image of an environment with finger touch gestures to make markers on it to benefit teleoperation and teleinteraction.

The first two tasks are essential to the TIUI for a remote operator to perform telepresence interaction by touching live video images. The third is an enhancing function, which can embed user's intention (such as grouping and cooperating) as well as user's knowledge (such as passage way and obstacles) into the system. Typical examples of on-line marking are to label obstacles and draw road edges for navigation.

**5.2 Finger Touch Gestures**

Following the expectation functions of finger-touch gestures mentioned above, we designed two classes of simple finger-touch gestures: one-finger touch gestures and two-finger touch gestures, following the work [Wroblewski 2010]. The one-finger touch gestures can be only used to interact with objects and environments, and the two-finger-touch gestures can be only used to control the telepresence robot. All gestures are intuitive and natural, and can be easily understood and performed by users, especially novice users.

*5.2.1 One-Finger Touch Gesture.* In our daily life, we can use our one-finger (such as forefinger) to accomplish almost all operations on panels or interfaces because they are composed of switches, buttons, and sliders. As for a joystick, it can be regarded as multiple buttons, arrow keys, or a track-point (of a ThinkPad). Therefore, we will design one-finger touch gestures for the TIUI to perform the teleoperation of most common objects by touching their live video images. The one-finger-touch gestures can be used to touch the live video image for

(1) selecting a tele-interactive object to be recognized,
(2) interacting with tele-interactive objects, and
(3) making markers on the environment.

Figure 6 shows the four one-finger touch gestures used in our system, including tap, long press, lasso, and drag for different interaction tasks. When an operator taps any region of a live video image, the system responds with the recognition result from computer vision algorithms or 2D barcode technologies. An operator can also lasso the region of an object in the image to assist the system to execute image segmentation for object recognition. An operator can drag an object image with one-finger to perform the sliding motion of the object, such as drawing a curtain, adjusting voice volume, moving a chess pawn, etc.

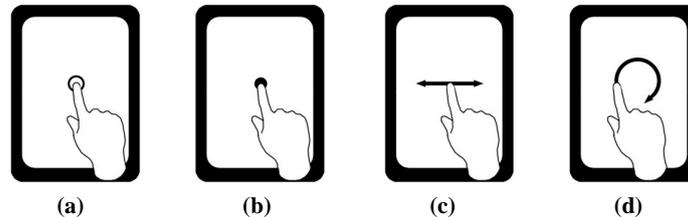

(a) (b) (c) (d)

Fig. 6. One-finger touch gestures for interaction with tele-interactive objects, including (a) tap, (b) press, (c) drag, and (d) lasso.

*5.2.2 Two-Finger Touch Gestures.* Figure 7 illustrates the two-finger touch gestures for teleoperation of the telepresence robot. An FDF live video image on the TIUI can be naturally divided into the two parts: the upper part and the lower part, according to the FF view and DF view. The upper part from the FF camera is characterized by the concentration of objects to be interacted with, and the lower part from the DF camera is more about the ground for navigation. The telepresence robot has two distinct parts: robot head and mobile robot base. Thus, we can easily associate the two parts of the FDF image with the two parts of the robot, respectively, to facilitate the interpretation of finger touch gestures. Concretely, when we touch the upper part of the TIUI with two-finger touch gestures, it means to control the robot head; when we touch the lower part of the TIUI with two-finger touch gestures, it means to drive the robot.

The idea of designing two-finger touch gestures for piloting the telepresence robot is inspired by the observation of human's boating and skating strokes. The two-finger touch gestures are defined the same as the human skating strokes, such as moving forward/backward or turning left/right. It is obvious that imitating skating strokes cannot achieve more efficient mobility than automatic driving robot by joysticks or graphic buttons. In fact, we are walking slower than driving a powered vehicle. But we argue that telepresence robot is indeed an embodiment of a remote operator in the local space, and should perform activities as the same as the local users, including walking, accompanying, and assistive action.

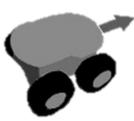

(a)

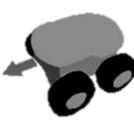

(b)

Fig. 7. Two-finger touch gestures to teleoperate the telepresence robot. (a) Two-finger touch gestures to pilot the mobile robot base, and (b) Two-finger touch gestures to control the robot head.

**6. SOFTWARE COMPONENTS OF THE SYSTEM**

The software of the system has to address the following issues:

(1) What does an operator intend to do by using finger touch gestures on the TIUI?
(2) Who in the live video image is being interacted with?
(3) How can an operator make markers in the live video image of a local environment for benefiting telepresence interaction?
(4) How can the system perform teleoperation and teleinteraction?

We will solve these four issues by computing the four modules: touch, recognition, knowledge, and control, as shown in Figure 8. Each module is computed independently, and can be regarded as one state of the system. Theoretically, the system could transit from any one state to another. In the telepresence interaction system, a remote operator has to join the loop of the system states via the TIUI, and plays a dominant role during execution according to the strategy of the Human-centered interaction operation [Dorais et al. 1999]. This means that an operator is watching the live video of the local environment in a remote space via the TIUI and looking for an object to be interacted with, and then touching its image with the finger touch gesture corresponding to an interaction task. So the touch module is a start state throughout telepresence interaction, and an operator can see the results computed by recognition, knowledge, and control module through the live video (i.e. visual feedback) to decide the next finger touch gesture.

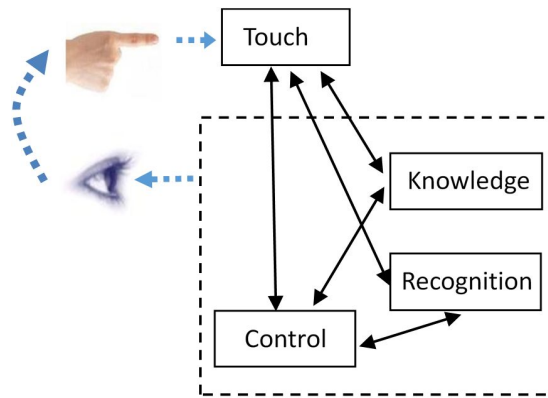

Fig. 8. Computing modules

### 6.1 Touch Module

The touch module is in charge of the detection and interpretation of finger touch gestures. The module needs to use the touchscreen drivers and I/O functions of a mobile device, and is completely executed on the mobile device. In our system, we simply define one-finger touch gestures and two-finger touch gestures on a pad for teleoperation (see Figure 6 and 7). The detection of finger touch gestures is used to obtain touching positions and tracks in the live video in order to tell which object is touched for teleinteraction or which region is touched for making markers. The interpretation of finger touch gestures refers to deriving or understanding which finger touch gesture is used by the operator.

According to the result of finger-touch gesture understanding, the system transits from the touch module to the recognition module, knowledge module, or control module. An operator looks at the live video on the TIUI as visual feedback to decide the next figure touch gesture for telepresence interaction.

### 6.2 Recognition Module

Recognition is crucial to our system as it can build the one-to-one correspondence between an object and its image in complex environments with multiple objects. In order to get the identification data of an object, we can use two recognition technologies: the computer vision (CV) and the tag scanning (2D barcode and RFID tags). The CV can learn or extract object features (such as corners, color, texture, shape, and appearance) for object classification and localization. It is the most universal and compact approach to do object recognition and image understanding, and significantly rapid growing in the last decades, but still suffering from the reliability and robustness in practical applications. If the CV is hard or fails to recognize an object, we can use the 2D barcode or RFID tags, two of the most popular and reliable tools for identification in our daily life. In our system, we adopt both the CV and the 2D barcode technologies to recognize objects.

Once an object is recognized, it is automatically locked on and tracked using the tracking methods [Wu et al. 2014, 2015] to ensure the continuity of teleoperation and teleinteraction. If the tracked object is lost, the system will return to the touch module at once and wait for operator's finger touch gesture to re-select the object.

### 6.3 Knowledge Module

Since knowledge acquisition involves complex cognitive processes (perception, communication, and reasoning), it is often a major bottleneck in the production of AI systems [Kidd 2012]. Fortunately, the loop of telepresence interaction contains the human operator link, which makes it easy to transfer operator's knowledge to the system. The task of the knowledge module is to embed operator's knowledge into the system for enhancing safety and efficiency of telepresence interaction.

In our system, an operator can use the TIUI to make markers on the live video image of a local environment to realize embedment of operator's knowledge into the system to facilitate the system performance with high efficiency, such as obstacle markers, route markers, door markers, and the like. Similar to the recognition module, a marked object is automatically locked on and tracked when the object is moving.

**6.4 Control Module**

Control module is to transform operator's intentions into control commands to realize teleoperation and teleinteraction. The operator's intention can be represented as finger touch gestures as we described in Section 5. Each tele-interactive object has its own control command set which is sent from the TIUI in a remote space, and can actuate the object to carry out a specific functions. So we only need to build the relationship between the finger touch gestures and the control commands of tele-interactive objects. When a touched object is recognized and locked on (by the recognition module), it becomes a slave to the remote operator and will be controlled through corresponding finger touch gestures. In our system, two-finger touch gestures are specifically designed to control the telepresence robot, thus, a remote operator can use two-finger touch gestures to drive the telepresence robot without recognition module. While there are a variety of tele-interactive objects in our daily life, the system has to recognize the touched object and then sends it the corresponding control command.

In order to ensure safety, we prefer to adopt a step-by-step action strategy based on supervisory control, which is strictly complied with the human-centered control framework. In other words, a step action corresponds to one of the finger touch gestures and the execution of each step-action is supervised by the operator via the TIUI. As a concrete example, an execution of the moving forward command of the telepresence robot only enables the robot to move forward a step (we can define a step as 100mm), and a continued moving forward needs a continued execution of the command under the operator's supervision.

**6.5 Basic State Transitions**

We use the numbers 0, 1, 2, and 3 to denote the computing modules of touch, recognition, knowledge, and control, respectively, and thereby Figure 8 can be simply redrawn as shown in Figure 9. In our system, three basic state transitions are abstracted to show the process mechanism of telepresence interaction. Figure 9a illustrates the first basic state transition, from State 0 to State 1, where a remote operator is watching the live video of an object as visual feedback (dashed blue arrow) while using the user interface to control the device. It is a typical conventional teleoperation strategy, a one-to-one remote control strategy only designed for teleoperation of one specific device, but it is not flexible to control other devices or multiple devices. The second basic state transition is designed for the teleoperation of multiple objects by one operator owing to the recognition module, as shown in Figure 9b. The transition from State 0 to State 2 is a distinct feature of our system compared with traditional teleoperation systems, which makes it feasible to interact with multiple common objects in complex environments. Figure 9c depicts the transition from State 0 to State 3, which is the third basic transition, an additional feature to embedding operators' knowledge into the system.

As we have seen, all three basic transitions start with State 0 (touch), and an operator gets the visual feedback from the TIUI about the results of State 1 (control), State 2 (recognition), or State 3 (knowledge). All three basic transitions are followed by operator's visual feedback to establish closed-loops of teleoperation and teleinteraction, and we call them closed-loop control, closed loop recognition, and closed-loop knowledge, respectively. In our system, the closed-loop knowledge and the closed-loop recognition are often carried out first, and then followed by the closed-loop control for exactly performing the telepresence interaction. As the telepresence robot is the only one robot as our embodiment, the robot control only adopts the one-to-one remote control strategy, i.e. closed-loop control strategy.

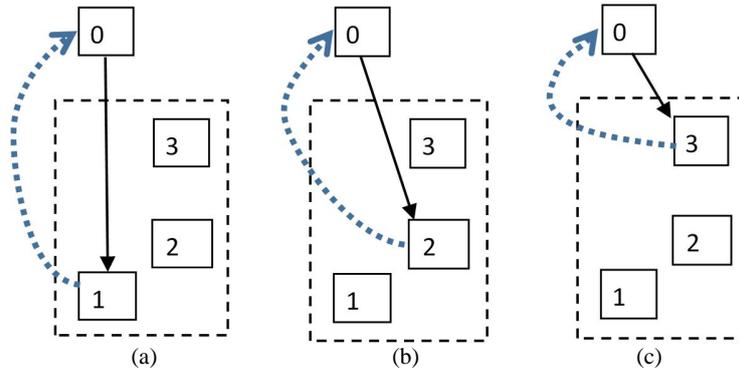

Fig. 9. Three basic state transition diagrams based on the four computing modules for telepresence interaction. 0-Touch, 1-Control, 2-Knowledge, and 3-Recognition. (a) Closed-loop control, (b) Closed-loop recognition, and (c) Closed-loop knowledge.

## 7. EVALUATION

We conducted a between-participants controlled laboratory experiment to evaluate the general usability of telepresence interaction by touching live video images. The experiment includes two tasks: one is to remotely drive the Mcisbot robot, and the other is to remotely interact with local objects. The goal is to gather user feedback for a remote operator to use the TIUI of a pad to interact with local objects while driving a telepresence robot.

We recruited 24 participants, 12 females and 12 males, from the local university to volunteer for our study, whose ages range from 18 to 25 years (M=23.525, SD=5.236). All of them use computers in their daily life, but they had no experience in the robot operation. Participants reported how familiar they were with the surfaces (e.g. Pads and Tablets) on a seven-point scale ranging from "1=not at all familiar" to "7 = very familiar" (M = 3.525 and SD = 1.523). Each participant was paid $10 per hour for participating in the experiment.

We built an experimental room in our lab simulating a living environment, containing sofas, tables, chairs, and some tele-interactive objects (see Figure 2). The room is specified as the local space. The room furnishings offer reference points for a remote participant to orient themselves within the environment. Participants located in another room, specified as the remote space, can use the TIUI on a pad to remotely operate the robot and interact with objects by touching their live video images with finger touch gestures.

### 7.1 Driving Task

It is an essential task for an operator to drive a telepresence robot in a remote space for telepresence interaction. Remotely operating the robot is a challenge task due to its high degree of operation and mobility. We evaluated the usability of the TIUI by comparing with the touchscreen based GUI because the latter is still a popular and advanced user interface in current systems [Micire et al. 2011; Tsui et al. 2015].

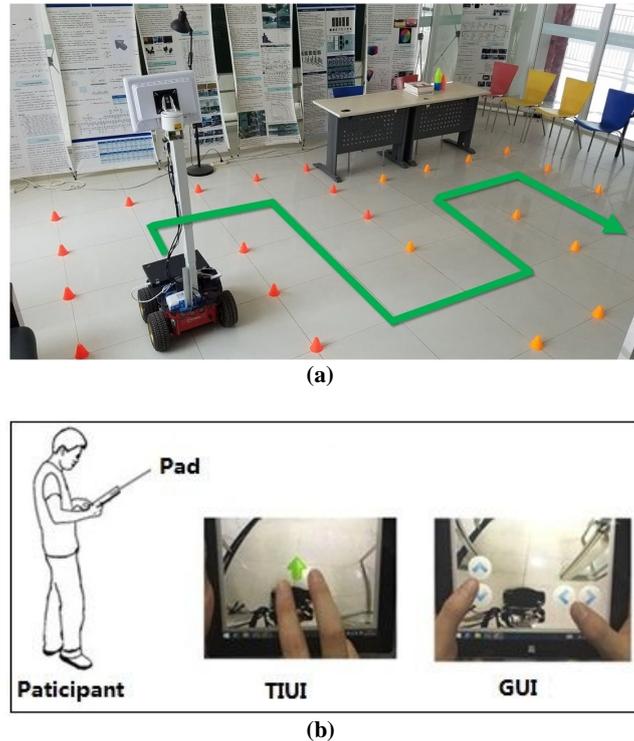

Fig. 10. Illustration of our experimental environment. (a) The physical arrangement of the local space. (b) A participant to pilot the Mcisbot robot in the remote space by using the TIUI and the GUI, respectively.

Figure 10 illustrates the experimental setup – a labyrinth-like obstacle course lined with colorful cups – for driving task. The driving experiment contained two sessions. The first session was that a remote participant drove the Mcisbot robot through the obstacle course by using finger-touch gestures (see Figure 7) on the TIUI of a pad. Following the first session, the second session asked a remote participant to drive the Mcisbot robot through the same obstacle course by using the graphic buttons on the GUI of the same pad.

We provided all participants with the instructions on how to use the TIUI and the GUI to drive the Mcisbot robot. In preparation for going through the obstacle course, participants were allowed to practice driving as many times as they wished, and the number of practice rounds each participant completed respectively with the TIUI and the GUI was recorded. Once participants indicated they were ready for the task, they made their way to the starting line and completed one loop of the obstacle course with the TIUI, and then proceeded to complete the other loop with the GUI. We recorded the time that participants spent and the number of cups they accidentally hit during driving the robot in one loop.

After finishing driving tasks, each participant was asked to fill in a questionnaire about evaluation with the maneuverability and perceived task success by experiencing operations with the TIUI and the GUI, respectively. For example, the perceived task success was measured via participant agreement with five items on a five-point scale (1 = strongly disagree to 5 = strongly agree), e.g. "I am confident that I used less time and hit fewer cups using the TIUI".

### 7.2 Interacting Task

As mentioned above, the TIUI allows a remote operator to drive the telepresence robot as well as interact with local objects just by touching their live video images. For evaluating the usability of interaction with local objects, we made a scenario for visiting and assisting a disabled elder who lived in a smart home (see Figure 3). The disabled elder has a limited capability of controlling the wheelchair, so extra assistance is needed. In this work, one of the authors took the role of the disabled elder.

We divided the interacting experiment into two sessions. In the first session, a participant visited the elder's home in person, and in the second session, the participant used the telepresence robot to visit the elder's home. Participants were instructed about the visiting procedure before performing the task. We asked the participants strictly following the procedure so that the visits of different participants were as consistent as possible.

Below is the visiting procedure of participants for the first session:
(1) A participant enters the room and sees the elder sitting in the sofa. The participant and the elder exchange greetings.
(2) The participant asks the elder to move from the sofa to the wheelchair (E) nearby, and then pushes him to the front of the curtain (D).
(3) The participant draws the curtain for the elder to look out the window, and they talk about the weather.

After completing the first session, the participant moved from the elder home (local space) to the other room (acting as a remote space) to carry out the second session. In the second session, the participant was instructed to use the Mcisbot robot through the TIUI on a pad to do the same visit as the first session:

(1) A participant uses the TIUI to drive the Mcisbot robot (C) to the elder's room and watches the elder sitting in the sofa. The participant and the elder exchange greetings.
(2) The participant asks the elder to move from the sofa to the wheelchair (E), and then uses the TIUI to push him to the front of the curtain (D).
(3) The participant lassoes the live video image of the curtain with the one-finger touch gesture, and then drags the curtain image for opening the curtain. The curtain is gradually opening as the finger dragging on the TIUI, and the elder looks out of the window. They talk about the weather.

When the participant completed the procedure of the second session list above, he/she was asked to fill in a questionnaire at once.

The questionnaire assessed the perceived spatial presence and measured the utility of the telepresence interaction by touching live video images on the TIUI. Participants were asked like the following question: "Please indicate your agreement with the word I drove the robot and interacted with those devices in the remote space as if I really walked and operated being there." Then participants rated their attitudes on a five-point scale, ranging from 1 (strongly disagree) to 5 (strongly agree).

### 7.3 Measures and Analysis

We investigated objective, subjective, and behavioral measures to evaluate the design of telepresence interaction system [Johnson et al. 2015].

The objective measures include the maneuverability, task performance, and situation awareness during driving by using the TIUI and GUI, respectively, in the remote space. To obtain the measures, we used the number of practice rounds and task completion time—timed from when the participant touched the touchscreen to when the participant indicated task completion. We recorded the number of bumping into cups as a measure of situation awareness.

The subjective measures sought to capture the maneuverability, perceived task success, utility of the teleinteraction approach, perceived difficulty of using the TIUI and the GUI, respectively. We administered two post-task questionnaires consisting of 16 questions, designed to capture these measures.

The behavioral measures refer to scores of how participants felt enjoyable during the telepresence interaction, and the scores were calculated by rating participants' face images on the telepresence robot screen. Independent coders (blind to the experimental conditions) rated each participant in terms of the quality of the participant's on-screen expression on a five-point scale (1 =strongly confused to 5 = strongly excited).

We used an analysis of variance (ANOVA) for testing the effects of different conditions upon each of the dependent variables (such as completion time, the number of bumps) to evaluate the system via the TIUI. For tests of statistical significance, we used a cut-off value of $p<0.05$.

### 7.4 Results

Figure 11 shows the results of the objective measures and subjective measures of the participants' teleoperation using the TIUI and the GUI, respectively. We can see that the participants practiced a few more rounds when using the TIUI (M=1.50, SD=0.511) than using the GUI (M=1.25, SD=0.442), but not significantly, $F(1, 22)=3.286$, $p=0.076$. Participants achieved better performance of completing the tasks using the TIUI than the GUI, and the average times of completing tasks were 102.25 seconds (SD = 30.087) and 122.29 seconds (SD = 32.378), respectively. We found a main effect of the TIUI on the task completion time, $F(1, 22) = 4.935$, $p = 0.031$. We also found a significant effect upon bumps, $F(1,22)=4.373$, $p<0.05$.

Participants reported that they experienced good maneuverability (M=4.38, SD=0.576) when using the TIUI as almost the same as when using the GUI (M=4.08, SD=0.881), $F(1, 22)=1.845$, $p=0.181$, but perceived better task success, $F(1,22)=5.665$, $p=0.022$.

Expressions of participants on the telepresence robot screen looked better (M=4.25, SD=0.671) when they used the TIUI than the GUI (M=3.58, SD=0.425), $F(1, 22) =5.423$, $p=0.024$.

Figure 12 shows that the participants responded very positively to the telepresence interaction system. Many participants also found the highly efficient teleinteraction metaphor by touching live video images of tele-interactive devices (M=4.45, SD=0.550). Most of the participants preferred using the TIUI than the GUI; 54% had no problems using the system after only one minute and 95% deemed that the TIUI is user-friendly and easy to use (M=4.50, SD=0.590). Most of the participants also experienced highly utility of the telepresence interaction via the TIUI and they really had "feelings of being there" (M=4.25, SD=0.532).

The results also show that the TIUI improved task performance over the GUI. We found differences in perceived task success and on-screen expression between finger touch gestures and graphic buttons, as finger touch gestures were designed to imitate the motion action of boating and skating strokes in our daily life. The TIUI made participants feel natural and intuitive, so they revealed much more enjoyable when using the TIUI with various finger touch gestures for teleinteraction and teleoperation.

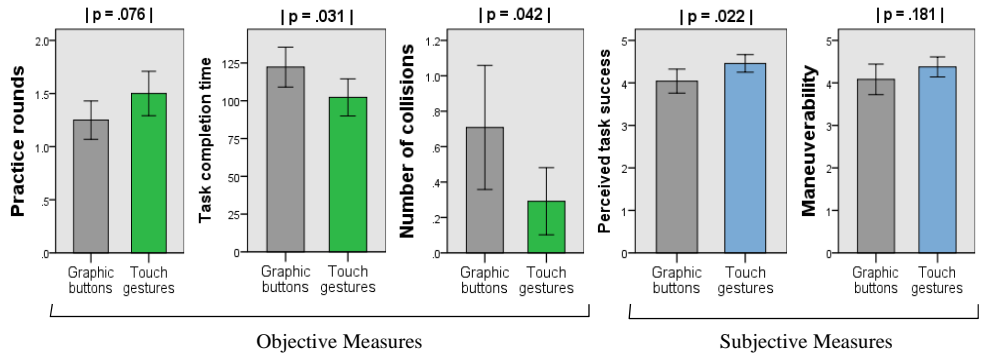

Fig. 11. Mean practice rounds, task completion time, the number of collisions, perceived task success, and maneuverability by using the TIUI and the GUI, respectively.

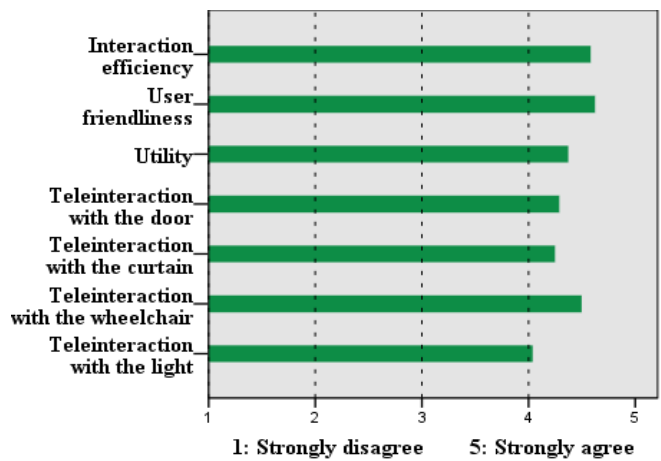

Fig. 12. Mean interaction efficiency, user-friendliness, utility, and user experience of telepresence interaction with the TIUI.

## 8. DEMONSTRATIONS AND DESIRABLE APPLICATONS

We have conducted three experimental demonstrations: teleoperation of the Mcisbot robot, telepresence interaction with a smart environment, and telepresence operation of a wheelchair robot. The implication of these demonstrations is that the telepresence interaction system with the TIUI is going to come into our daily lives. The system can provide a remote operator with the TIUI to have not just verbal and non-verbal communication but teleoperation and teleinteraction. The system can allow a remote operator to readily access the system anywhere by using the TIUI on a pad or smartphone to achieve being present in two places at the same time. These demonstrations have shown the potential and desirable applications, such as accompanying elderly parents, attending a conference, and collaborating on a research project.

**8.1 Teleoperation of the Mcisbot Robot**

Affordable telepresence robots have started to become commercially available, and the pace of research and development has accelerated accordingly [Herring et al. 2016]. Different from the existing telepresence robots, we use a novel user interface, the TIUI, to pilot the telepresence robot. To our best knowledge, this is the first system to drive a telepresence robot in a remote environment by directly touching its live video image with finger touch gestures.

In this section, we demonstrate how a remote operator uses the TIUI on a pad to pilot the Mcisbot robot efficiently with finger touch gestures. Since the FDF live video of the TIUI is the first person view of the Mcisbot robot, the mobile robot base can be constantly located at the lower part of the FDF live video image. A two-finger touch gesture detected at the lower part of a live video image means just to control the mobile robot base to move forward/backward and/or turn left/right according to the gesture definitions (see Figure 7a), and a two-finger touch gesture detected at the upper part is just used to control the robot head to look around (see Figure 7b). Figure 13 shows a remote operator uses the two-finger touch gestures at the lower part of the live video image (i.e., the image of the mobile robot base) to control the robot moving forward, turning left/right. The moving speed depends on the frequency of using touch gestures, similar to skating strokes.

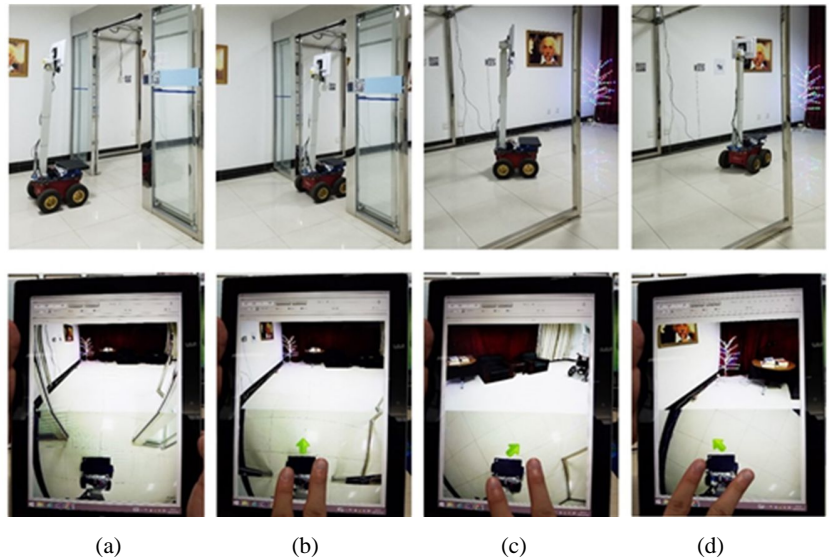

(a)          (b)          (c)          (d)

Fig. 13. Two-finger touch gestures performed at the lower part of the TIUI to teleoperate the mobile robot base of the Mcisbot. Bottom: the TIUI of a pad used by an operator in a remote space. Top: the corresponding movements of the Mcisbot robot in the local environment, controlled by the remote operator with the two-finger gestures, (a) Reference position, (b) Move forward, (c) Turn left, and (d) Turn right.

Figure 14 shows a remote operator uses the two-finger touch gestures at the upper part of the live video image (e.g. the image of the robot head) to control the robot head to move up and down for a proper height and look forward, left, and right Figure 14a shows the starting state of the Mcisbot robot with a height of 1200mm to look forward. Figure 14b shows a remote operator uses the two-finger touch gesture at the upper part of the live video image to control the vertical lifting post to move up to look forward. Figures 14c and 14d illustrate that a remote operator controls the pan-tilt platform to perform looking left and right actions.

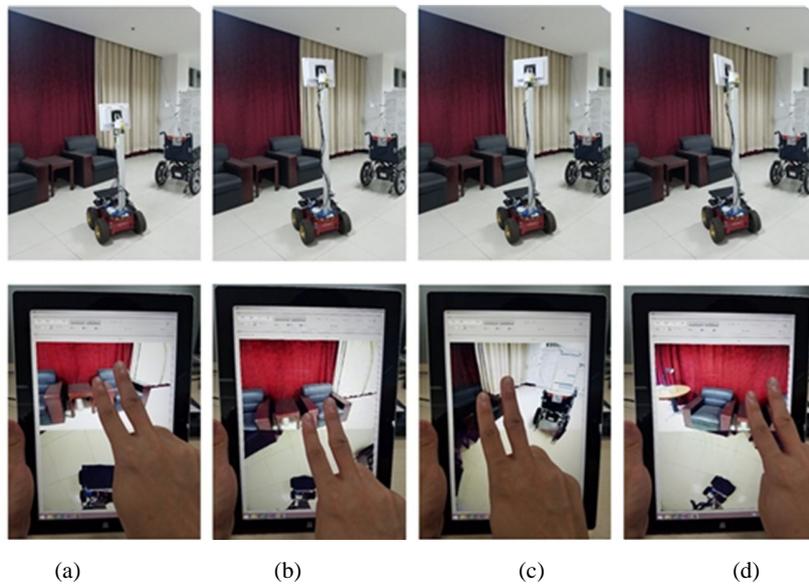

(a)          (b)          (c)          (d)

Fig. 14. Two-finger touch gestures performed at the upper part of the TIUI to teleoperate the Mcisbot head. (a)Look forward in the height of 1200mm, (b) Look forward in the height of 1750mm, (c) Look left, and (d) Look right.

**8.2 Telepresence Interaction with Smart Environments**

Comparing with the conventional teleinteraction, the telepresence interaction via the TIUI includes the verbal and non-verbal communication, teleoperation, and teleinteraction. A remote operator uses the TIUI to not only communicate with the persons of a local space, but also operate and interact with the tele-interactive objects of the local space by touching their live video images. Following the teleoperation of the telepresence robot described in Section 8.1, we tried to test our system to perform telepresence interaction with the smart environment we built in our lab (see Figure 3).

Figure 15 shows that a remote operator uses the TIUI to drive the Mcisbot robot to the front of the auto-door and find password access control panel to open the door. The operator can lasso the image of the 2D barcode beside the panel with the one-finger touch gesture to recognize the panel, and the recognition result includes the localization of the panel and the position of each button in the image. Then the operator can tap the button image to input the password to open the door. Figure 15a shows the door closed before teleoperation and Figure 15d the door opened after teleoperation. This demonstration implies that the system can support teleoperation and teleinteraction of any buttons or keys in an intuitive and natural way.

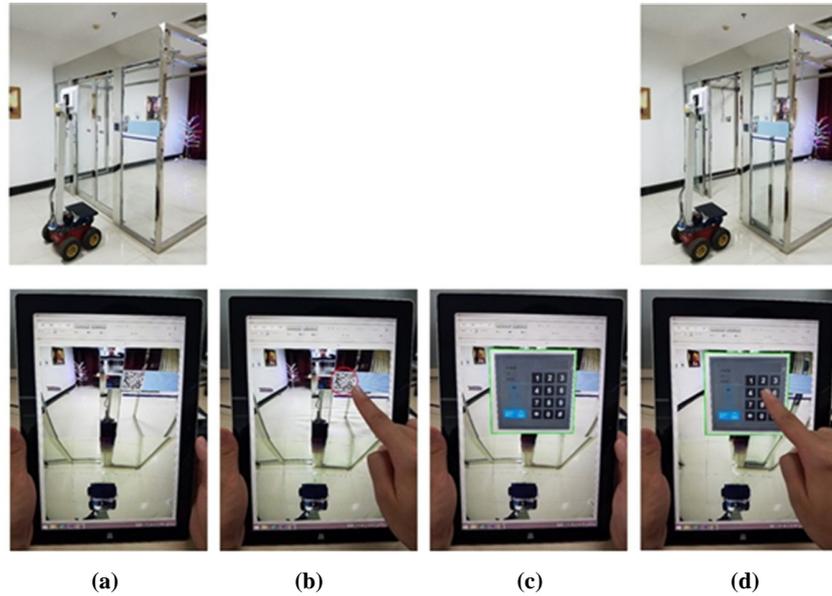

**(a)** **(b)** **(c)** **(d)**

Fig. 15. A remote operator uses the TIUI to open the auto-door with password access control. (a) A remote operator uses the Mcisbot to look for the password panel (the door is closed). (b) The operator lassos the image of the 2D barcode (red circle). (c) The system recognizes and localizes the password panel (green box). (d) The operator taps the button image to input password to open the door (the door is opening).

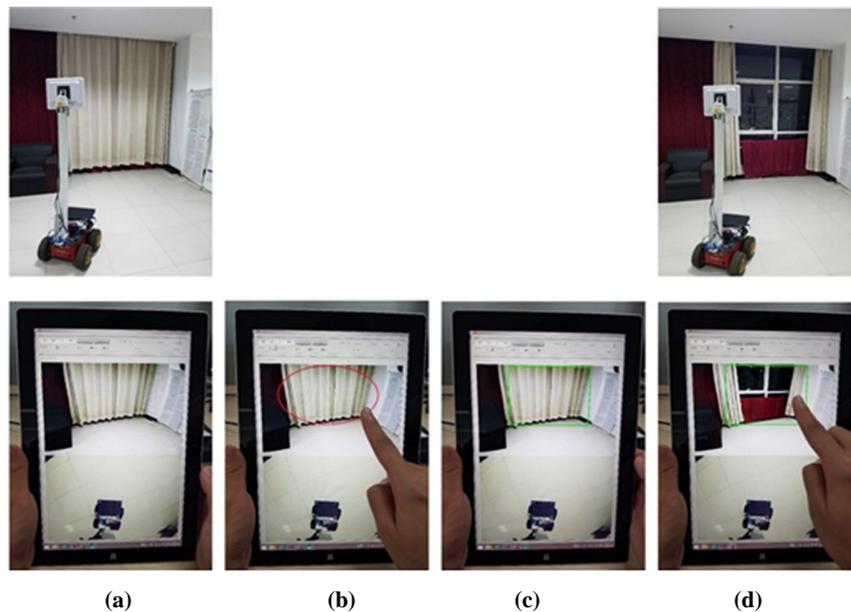

**(a)** **(b)** **(c)** **(d)**

Fig. 16. A remote operator uses the TIUI to draw the curtain. (a) The operator remotely driving the Mcisbot and looking for the curtain. (b) The operator lassos the curtain image for image segmentation (red circle). (c) The system is guided by the result of image segmentation to recognize and localize the curtain (green box). (d) Following the result of recognition, the operator uses the TIUI to draw the curtain with the one-finger touch gesture.

Figure 16 shows that an operator can easily use the TIUI to remotely draw the curtain as if he/she did in the local space. The operator uses the Mcisbot robot to look for the curtain (Figure 16a) and lasso the curtain image with the one-finger touch gesture for image segmentation. The image segmentation result guides the system to recognize and localize the curtain (Figures 16b and 16c), and then the operator can draw the curtain as if he/she did in the local space (Figure 16d).

**8.3 Telepresence Operation of a Wheelchair**

One of the important application domains for telepresence robots is elderly care and health care, where telepresence robots can be profitable and contribute to the prevention of problems related to loneliness and social isolation [Coradeschi et al. 2011; Cesta et al. 2016]. Telepresence robots are emerging to allow elderly people at home to communicate with the family members or caregivers, and it is found that telepresence robots enable elderly people to regard the telepresence robot as a representation of the robot operator (family members) [Tsai et al. 2007]. Nowadays, commercially available telepresence robots include Giraff robots and VGo robots, designed specifically for elderly and disabled people. But all of these telepresence robots are insufficient in teleinteraction and teleoperation for achieving telepresence interaction with the environment. For example, one can remotely drive a telepresence robot in his/her aged parents' home and chat with them to relieve the elderly loneliness, but he/she cannot operate the elder's wheelchairs which are their mobility support devices to get around.

The Mcisbot robot is a telepresence robot which can be a remote operator' embodiment to accompany the elderly people and assist them to perform some necessary tasks, such as opening the door, drawing the curtain, and turning on/off lighting, as described in Section 8.2. In this section, we demonstrate how a remote operator uses the TIUI to push a WiFi wheelchair as he/she did in the local space, e.g. one can visit the elder home in person and push the wheelchair for the elder to move around (Figure 17a), and one can also use the Mcisbot robot as his embodiment to remotely do the same task (Figure 17b).

Figure 18 illustrates a powered wheelchair with WiFi communication in a local space (such as houses or nursing houses) which is pushed by a remote operator via the Mcisbot robot. First, a remote operator uses the TIUI to drive the Mcisbot to the back of the wheelchair (Figure 18a). Second, the operator touches the live video image of the wheelchair in the TIUI with the one-finger touch gesture to recognize and localize the wheelchair (2D barcode), and then the wheelchair image is marked by a bounding box and ready to be pushed (Figure 18a). Third, the operator touches the image of the wheelchair with the one-finger touch gesture to push the wheelchair for moving forward (Figure 18b), turning right (Figure 18c), and turning left (Figure 18d).

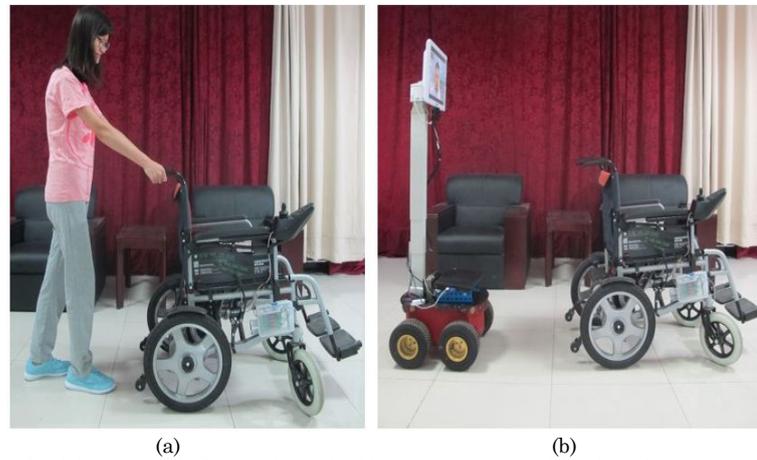

(a) (b)
Fig. 17. A wheelchair for elderly people pushed by a person (a) and by the telepresence robot (b).

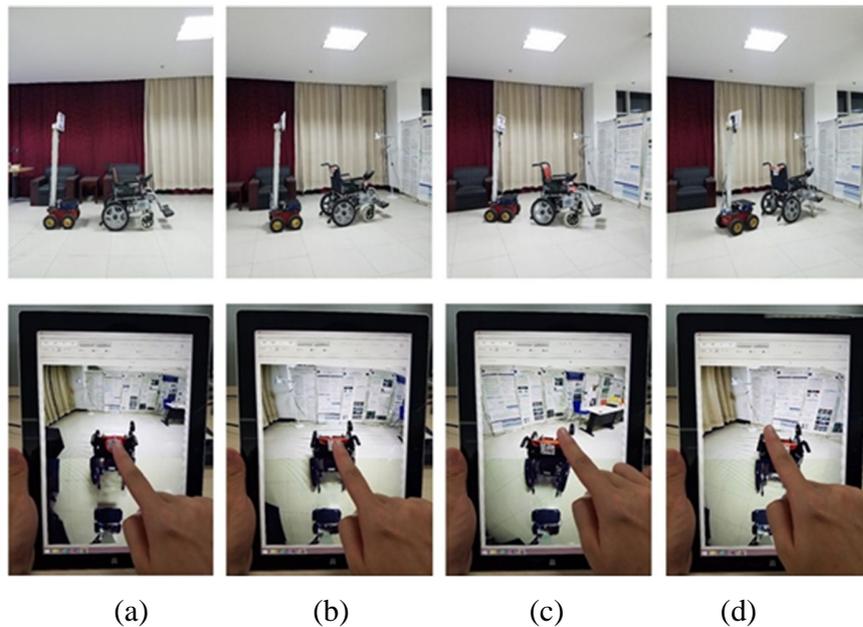

(a) (b) (c) (d)
Fig. 18. The remote operator pushes the wheelchair through the TIUI with one-finger gesture, and the telepresence robot moves to follow the wheelchair.

## 9. DISCUSSIONS

We have developed a telepresence interaction system with a touchable live video image based user interface. The system enables a remote operator to not just drive the telepresence robot and video-chat with the local person(s), but also operate and interact with the local objects. However, we thought of many factors which will slow down the system growth.

## 9.1 Smart Environment

In smart environments, everyday devices are smart by embedding processors, sensors and actuators, and connecting to the internet [Weiser 1991]. For our system to get into use, we might face the following challenges.

*9.1.1 Smart Devices.* The telepresence interaction system is designed for smart environments with tele-interactive objects. However, most of everyday devices in our living or working environments currently are still powered electrically, and need to "add on" a tele-interactive box, similar to a set-top box, to make it smartlized. A tele-interactive box can be basically composed of a WiFi card for communication, a relay for actuation, and a 2D barcode tag covered on the box for identification, and it is very cheap as all these components are off-shelf products.

*9.1.2 Wireless Communication Networks.* Our system requires the broadband wireless communication infrastructure, such as WiFi, to transmit high-quality live videos for the TIUI. We successfully tested our system in the indoor environment with a 2.4G wireless router connected to the LAN of the building. Furthermore, new generation of fiber optic communication technologies and high-speed wireless access to the internet are accelerating [Yin et al. 2015], which will greatly enhance the broadband wireless communication infrastructure of both indoor and outdoor environments and benefit the development of telepresence interaction systems.

*9.1.3 Barrier-Free Paths.* We prefer to use wheeled-robots as the mobile base of telepresence robots because of their low cost, good maneuverability, high reliability, and high maintainability. In recent years, some simple prototypes (webcam on wheels and iPad on wheels) have received attention [Dahl and Boulos 2013; Kristofferson et al. 2013]. But wheeled-robots require barrier-free paths or zones throughout the area of their activities, including lifts, ramps, auto-doors, and all floors. Fortunately, these requirements are identical to the wheelchair-users and tactile lines for blind people. There has been a broader view on the total service or activity that is going to be universal, such as visiting a cinema, travelling by train, or being a tourist in a city [Steinfeld 2013].

## 9.2 Safety

The safety is one of the most concerning issues during tele-operating a mobile robot since it can produce powerful and rapid movement that might cause a hazard to humans surrounding it [Vasic and Billard 2013]. With respect to a telepresence robot controlled in a remote location, we considered two aspects to ensure the safety. First, we have made use of a step-by-step action strategy based on supervisory control, as we mentioned in Section 6.4. Each step-action only can produce limited movement to avoid causing the hazard to humans surrounding it. Second, our system adopts the wheeled-robot structure which is not equipped with arms and manipulators to dispel some fears concerning going out of control and endangering local persons or property, in contrast to existing systems [Adalgeirsson and Breazeal 2010; Koceski and Koceska 2016].

## 9.3 Acceptance

The acceptance of telepresence interaction by touching live video images, similar to the acceptance of telepresence robot technologies [Beer and Takayama 2011], is an open problem. In this work, we have taken into account the acceptance of telepresence interaction technology, according to the well-known technology acceptance model (TAM) [Davis 1986] which is affected by two factors: perceived usefulness and perceived ease-of-use.

*9.3.1 Perceived Usefulness.* Perceived usefulness refers to how much the user believes that the technology will help to improve the performance/efficiency [Davis 1989; Alanazi et al. 2015]. In the remote space, if an operator would like to drive the existing telepresence robots to its goal position, there must have somebody in the local space to assist them for doing some assistance tasks, as feeling disabled [Lee and takayama 2011]. The Mcisbot, a telepresence robot in our system, can go anywhere independently in the local space without any assistance of the local person. This is because an operator can remotely operate any objects via the TIUI, such as an open a door and operate an elevator. Moreover, the remote operator can help the local person, e.g. elderly people, to accomplish many tasks. So our system is of great potential usefulness in the smart world.

*9.3.2 Perceived Ease-of-Use.* The perceived ease-of-use refers to the extent to which the user is comfortable to use the features of the technology [Davis 1989; Alanazi et al. 2015]. In the design of our system, we considered three factors for perceived ease-of use: (1) a pad or smartphone which makes an operator ready to get access to the system anywhere, (2) the TIUI which allows an operator to remotely operate the robot and any other objects for telepresence interaction, including monitoring, smart housing, office settings, and (3) wide view field of live video for navigation. A remote operator uses the TIUI to drive the robot with a very wide view of the ground on which the robot moves for easy navigation and surrounding perception.

### 9.4 Limitations and Future Work

The first limitation of our study is that we only designed a small set of finger touch gestures to realize the telepresence interaction system by touching live video images. We will design more touch gestures to iteratively get better user experiences to satisfy the TAM [Davis 1986].

The second limitation is the system delay, including communication and computation. Currently, our system spent 100ms to obtain the live video, including capturing live videos from the FF camera and the DF camera, and stitching the two live video images into the FDF live video image. We will reselect hardware components to broaden data channels for communication, and redesign software components to speed up the computation.

The third limitation is that our system cannot use the autonomous strategy to carry out the teleoperation and teleinteraction in view of safety and cost. We have developed a semi-autonomous strategy only based on live video images for the robot navigation and obstacle avoidance [Shen et al. 2016]. We will continue to study semi-autonomous strategies and machine learning methods based on live video images to improve teleoperation efficiency and teleinteraction comfortableness.

The fourth limitation is that our study only recruited a small sample from a university campus, and we will generalize our study to a larger and more representative sample of users.

### 10. CONCLUSIONS

We have presented the telepresence interaction framework with touching live video images towards smart environments to realize not only verbal and non-verbal communication, but also teleoperation and teleinteraction. We designed a novel User Interface (TIUI) which allows a remote operator to drive a telepresence robot and interact with the tele-interactive objects by using a pad and finger touch gestures. We developed the telepresence interaction system composed of a telepresence robot and tele-interactive objects in a local space, the TIUI in a remote space, and the wireless communication network connecting the two spaces. The evaluation of the TIUI and the system demonstrations show that the proposed framework and methodology are promising and useful.